\documentclass[11pt]{article}
\usepackage[margin=1in]{geometry}
\usepackage{booktabs}
\usepackage{multirow}
\usepackage{amsmath}
\usepackage{graphicx}
\usepackage{enumitem}
\usepackage{xspace}
\usepackage{array}
\usepackage{makecell}
\usepackage{microtype}
\usepackage{placeins}
\usepackage[numbers,sort&compress]{natbib}
\usepackage[colorlinks=true,linkcolor=blue,citecolor=blue,urlcolor=blue]{hyperref}
\usepackage{caption}
\usepackage{subcaption}
\providecommand{\Description}[1]{}
\newcommand{\sys}{LHF\xspace}

\newcommand{\smalltab}{\scriptsize}
\title{Diagnosing and Mitigating Retrieval Bottlenecks in LLM-Based Cold-Start Recommendation}
\author{
\begin{tabular}{c}
Zhe Dong$^{1,*}$ \\
\texttt{zhe.dong@maine.edu} \\
\texttt{dongzhe181@gmail.com} \\
$^1$University of Maine at Presque Isle
\end{tabular}
\and
\begin{tabular}{c}
Fang Qin$^2$ \\
\texttt{fangq@stanford.edu} \\
$^2$Stanford University
\end{tabular}
\and
\begin{tabular}{c}
Manish Shah$^3$ \\
\texttt{shahmh@ieee.org} \\
$^3$Independent Researcher
\end{tabular}
\and
\begin{tabular}{c}
Yicheng Wang$^3$ \\
\texttt{ethanwang63@163.com} \\
$^3$Independent Researcher
\end{tabular}
}
\date{Preprint / extended version \\
$^*$Corresponding author}

\begin{document}
\maketitle
\begin{abstract}
Large language models (LLMs) are increasingly used as rerankers in recommender systems, with the expectation that semantic understanding will help in cold-start and long-tail regimes. We test this assumption with a five-domain benchmark that explicitly separates \emph{reranking quality} from \emph{retrieval coverage}. In a positive-controlled regime where the gold item is guaranteed present, calibrated LLM rerankers fail to consistently outperform strong collaborative and content baselines under natural traffic, and within-family scaling from Qwen3-8B to Qwen3-32B narrows but does not close the gap on most domains. In a retrieval-realistic regime where the gold item is not injected, the bottleneck is more severe: standard single retrievers place the gold item in a 200-item pool only 4.6--22.9\% of the time, largely because 32--91\% of cold-start targets are brand-new items with no training interactions. We introduce \sys, a validation-trained learned hybrid fusion layer over a multi-retriever union pool, as a retrieval-side realizability baseline. \sys is the only combiner we test that beats every single retriever on all five domains and recovers 17--61\% of oracle coverage headroom on content-rich domains, but only 5--7\% on collaboratively strong domains. End-to-end experiments reveal the remaining mismatch: learned non-LLM ranking exploits the \sys pool, while prompt-level LLM reranking often degrades it. LLMs exhibit pockets of semantic cold-start advantage, especially in text-rich domains when the item is already present, but this advantage is largely unreachable in current retrieve-then-rerank pipelines. We release the benchmark protocol, splits, prompts, evaluation tooling, and archived reproducibility artifacts: data at \url{https://doi.org/10.5281/zenodo.20991039} and code at \url{https://doi.org/10.5281/zenodo.20993306}.
\end{abstract}

\section{Introduction}
LLM-based recommendation commonly follows a multi-stage pipeline: a fast retriever proposes a few hundred candidates and an LLM reranks them using item text and user history. The appeal is intuitive. Cold-start users, new items, and long-tail content are precisely where interaction signals are sparse and semantic content should matter. Yet a reranker can only reorder items it sees. If the correct item never enters the candidate pool, increasing LLM size or refining prompts cannot recover it.

\begin{figure}[t]
    \centering
    \includegraphics[width=0.98\textwidth]{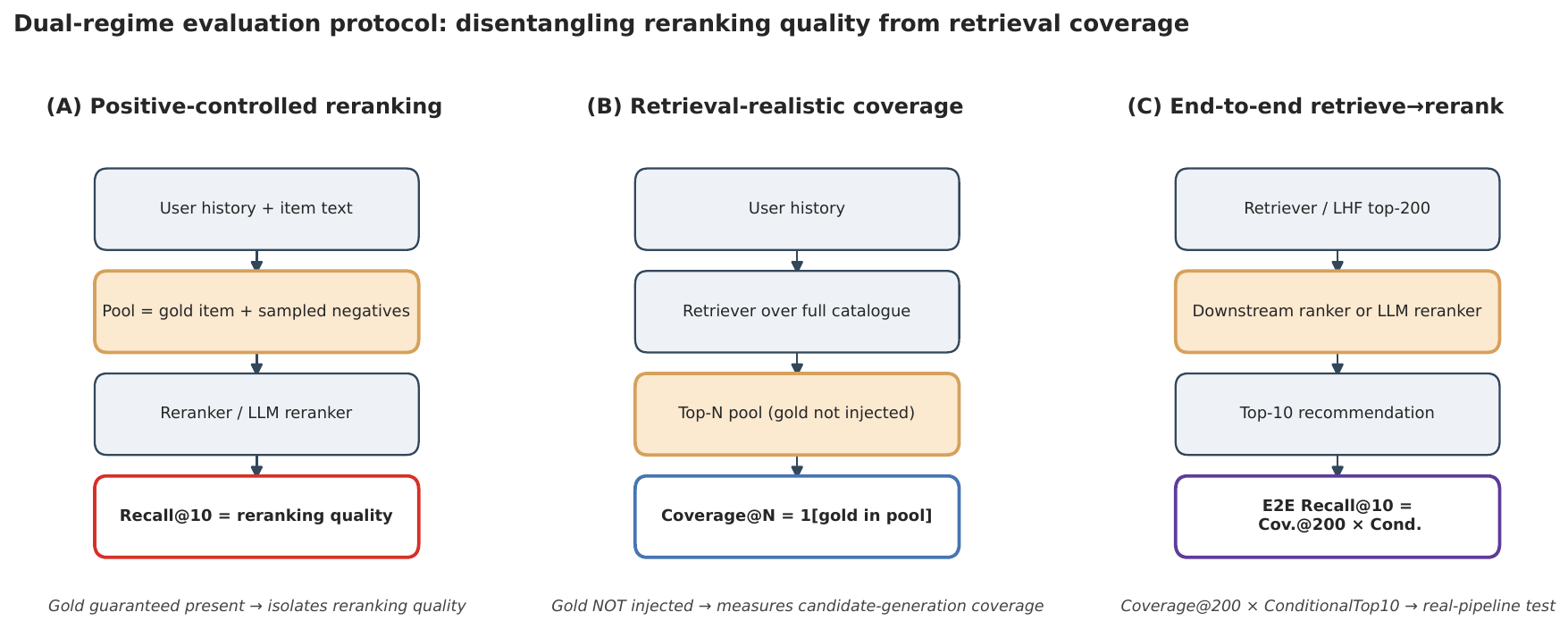}
    \Description{Three-panel schematic showing positive-controlled reranking, retrieval-realistic coverage, and end-to-end retrieve-then-rerank evaluation.}
    \caption{Dual-regime evaluation protocol. Positive-controlled pools isolate reranking by injecting the gold item. Retrieval-realistic pools remove this injection and measure coverage. End-to-end evaluation feeds the realistic pool to a downstream ranker.}
    \label{fig:protocol}
\end{figure}

We study that mismatch. Rather than evaluate LLM recommendation only in injected-candidate pools, we separate the problem into three regimes (Fig.~\ref{fig:protocol}): (i) a \emph{positive-controlled} pool, where the gold item is guaranteed present and the question is whether the ranker can identify it; (ii) a \emph{retrieval-realistic} pool, where each retriever searches the full catalogue and the question is whether the gold item appears at all; and (iii) an end-to-end retrieve-then-rerank pipeline, where recall decomposes as
\begin{equation}
\eet@10 = \cov@200 \times \cond@10,
\end{equation}
with $\cond@10$ the probability that a reranker puts the gold item in the top 10 given that it was covered.

Across Amazon Arts, Amazon Video Games, MIND news, MovieLens-20M, and Yelp restaurants, we find a consistent picture. LLM rerankers have genuine semantic cold-start ability in controlled pools, especially on news, but this advantage does not transfer to realistic pipelines. Standard retrievers cover the gold item in only 4.6--22.9\% of top-200 pools, and even strong dense encoders, a two-tower retriever, and a learned fusion raise this only to 6.1--24.3\%. Meanwhile, when coverage is non-trivial, prompt-level LLM reranking often demotes correct items that collaborative or learned rankers place near the top.

We contribute four main results: (1) a dual-regime and end-to-end benchmark that disentangles reranking from retrieval; (2) a five-domain diagnosis showing that within-family LLM scale narrows but does not remove the gap to strong collaborative baselines because retrieval coverage is capped by item-new targets; (3) \sys, a validation-trained retrieval-side realizability baseline that partially converts multi-retriever complementarity into coverage gains; and (4) evidence that the improved \sys pool is exploited by a lightweight learned ranker but not by prompt-level LLM reranking. We also provide a graph-prompt side-signal audit and cost characterization. The reproducibility artifacts are archived separately as a data release (\url{https://doi.org/10.5281/zenodo.20991039}) and a code release (\url{https://doi.org/10.5281/zenodo.20993306}); the code is also available at \url{https://github.com/dongzhe1/rec-diag-bench/tree/0.0.0}.

\section{Related Work}
\textbf{LLMs for recommendation and reranking.} Prior work frames recommendation as language modeling or instruction following, including P5~\cite{geng2022p5}, TALLRec~\cite{bao2023tallrec}, and zero-shot or instruction-tuned LLM rerankers such as LLMRank, RecRanker, and ChatGPT-style recommenders~\cite{hou2024llmrank,luo2024recranker,liu2023chatgptrec}. Surveys summarize LLM-for-recommendation pipelines and paradigms~\cite{wu2024llmrec,lin2025llmrecsurvey}. Much of this work evaluates in warm or candidate-injected settings. We stress-test LLM reranking under temporal cold-start and explicitly separate the reranking step from candidate generation.

\textbf{Cold-start and hybrid retrieval.} Cold-start has long motivated content-to-collaborative hybridization, including DropoutNet~\cite{volkovs2017dropoutnet}, Heater/MWUF~\cite{zhu2021mwuf}, MetaEmb~\cite{pan2019metaemb}, MeLU~\cite{lee2019melu}, contrastive cold-start learning~\cite{wei2021clcrec}, and modality-aware ranking~\cite{he2016vbpr}. Our work quantifies where content helps: not primarily in prompt-level reranking, but in retrieving new items that have no interaction history. We also compare standard dense retrieval, strong off-the-shelf encoders, a frozen two-tower retriever, and learned fusion against collaborative retrievers.

\textbf{Recommendation baselines and evaluation.} Our baseline pool covers matrix factorization~\cite{rendle2009bpr}, graph CF~\cite{wang2019ngcf,he2020lightgcn}, sequential recommendation~\cite{kang2018sasrec}, dense retrieval~\cite{reimers2019sbert,xiao2023bge,chen2024bgem3,wang2022e5,karpukhin2020dpr,yi2019sampling}, reciprocal-rank fusion~\cite{cormack2009rrf}, and gradient-boosted ranking~\cite{ke2017lightgbm}. Recommender evaluation is sensitive to negative sampling, leakage, and weak baselines~\cite{dacrema2019progress,dacrema2021troubling,canamares2020target}. We extend this perspective to LLM recommendation: candidate injection, stratified cold-tail subsets, negative sampling, and tail priors can all over-credit LLM rerankers.

\textbf{Closest prior work.} The closest prior diagnostic study is a single-domain cold-start reranking analysis on movies~\cite{lemdiasova2026coldrerank}. It links LLM/cross-encoder reranking failures to coverage and exposure. We differ by evaluating real instruct LLMs, using a positive-controlled regime to isolate reranking, adding full-catalogue retrieval and end-to-end tests across five domains, and introducing a learned retrieval-side mitigation.

\section{Setup and Protocol}
\subsection{Datasets and temporal buckets}
We evaluate five public domains: Amazon Reviews 2023 Arts and Video Games~\cite{hou2024blair,amazonreviews2023}, MIND small news~\cite{wu2020mind}, MovieLens-20M~\cite{harper2015movielens}, and Yelp Philadelphia restaurants~\cite{yelpdataset}. We treat ratings above a domain-specific threshold or clicks as positive feedback, split interactions chronologically into 70/10/20 train/validation/test, evaluate users seen in training, and cap test users at 5,000 where applicable. At retrieval time, the candidate catalogue includes items whose metadata is available before the request timestamp, while all post-training interactions are withheld; this is what makes text retrieval possible for \texttt{item\_new} targets while keeping interaction signals unavailable.

Cold-start labels are computed from the training split only. Each test interaction is assigned to one primary bucket by priority: \texttt{item\_new} (target has zero training interactions), \texttt{item\_cold} (bottom active-item frequency), \texttt{long\_tail}, \texttt{user\_cold}, and \texttt{warm}. Table~\ref{tab:data} shows that temporal evaluation is dominated by item-new targets.

\begin{table}[t]
\centering
\caption{Temporal test buckets. Item-new targets have zero training interactions and dominate several domains.}
\label{tab:data}
\smalltab
\begin{tabular}{lrrrrrr}
\toprule
Domain & New & Cold & Tail & U-cold & Warm & Total \\
\midrule
Arts & 1768 & 372 & 174 & 360 & 2326 & 5000 \\
VideoG & 3481 & 145 & 56 & 242 & 1076 & 5000 \\
MIND & 4528 & 57 & 4 & 40 & 371 & 5000 \\
ML-20M & 1894 & 3 & 22 & 61 & 971 & 2951 \\
Yelp & 817 & 125 & 89 & 231 & 1330 & 2592 \\
\bottomrule
\end{tabular}
\end{table}

\subsection{Models, prompts, and metrics}
Candidate generators include popularity, item-kNN, TF-IDF, Markov, BPR, LightGCN, SASRec, graph co-occurrence and embeddings, SBERT/BGE dense retrieval, and several fusions. Rerankers include a cross-encoder and prompt-based LLM rerankers. The primary LLM is Qwen3-8B~\cite{qwen3technical}; Qwen3-32B and Llama-3.3-70B-Instruct~\cite{llama33modelcard} are scale checks. Candidates are scored by the log-probability of a calibrated ``Yes'' answer. Unless noted, results use seed 42, top-200 retrieval, full history, and Recall@10 or Coverage@200.

\section{LLM Reranking in Controlled Pools}
\subsection{Gold-injected positive-controlled evaluation}
The positive-controlled pool contains the gold item and sampled negatives. It asks whether a ranker can identify a correct item after retrieval has already succeeded. Table~\ref{tab:rerank} shows representative Recall@10 rows. Qwen3-8B does not win any domain; LightGCN dominates ML-20M, graph/content fusion wins Arts/Yelp, and a cross-encoder wins Video Games.

\begin{table}[t]
\centering
\caption{Positive-controlled reranking Recall@10. Gold is injected, so this isolates reranking quality.}
\label{tab:rerank}
\smalltab
\begin{tabular}{lccccc}
\toprule
Model & Arts & VideoG & MIND & ML-20M & Yelp \\
\midrule
Popularity & .115 & .140 & .038 & .419 & .093 \\
Item-kNN & .098 & .096 & \textbf{.107} & .371 & .157 \\
LightGCN & .147 & .140 & .069 & \textbf{.490} & .159 \\
Fusion & \textbf{.220} & .157 & .084 & .390 & \textbf{.159} \\
Cross-enc. & .152 & \textbf{.163} & .100 & .129 & .085 \\
LLM-8B & .145 & .143 & .092 & .219 & .085 \\
\bottomrule
\end{tabular}
\end{table}

Under natural traffic, scaling narrows but does not close the gap. Qwen3-32B improves over 8B on four domains, but still loses to the best collaborative or fusion baseline on four of five domains (Table~\ref{tab:scale}). The exception is MIND, where interaction signal is weak and text is rich.

\begin{table}[t]
\centering
\caption{Natural-reweighted reranking Recall@10 and scale. Scaling helps, but does not rescue most domains.}
\label{tab:scale}
\smalltab
\begin{tabular}{lcccc}
\toprule
Domain & Best CF/fusion & LLM-8B & LLM-32B & LLM-70B \\
\midrule
Arts & \textbf{.252} & .094 & .141 & .118 \\
VideoG & \textbf{.233} & .155 & .211 & .172 \\
MIND & .093 & .089 & \textbf{.169} & .137 \\
ML-20M & \textbf{.602} & .210 & .190 & .116 \\
Yelp & \textbf{.274} & .061 & .089 & .050 \\
\bottomrule
\end{tabular}
\end{table}

\subsection{Negative sampling and prompting checks}
Evaluation choices can flip controlled-pool conclusions. With uniform negatives, CF appears stronger on Video Games; with popularity-weighted negatives, CF/popularity methods rank popular distractors above the gold item and collapse, while the LLM is more robust (Video Games: best-CF .157 to .084, LLM .143 to .171). On Yelp, the CF-LLM gap shrinks from .074 to .010. In all controlled-pool comparisons, every ranker sees the same candidate pool for a user; negatives are sampled from non-positive items after excluding the user's observed history and held-out positives, either uniformly or proportional to training-set item frequency. Prompting variants do not rescue the LLM: pointwise yes/no log-probability beats 0--100 scoring and listwise top-10 selection on the tested subsets (VideoG .040 vs .020/.020; MIND .047 vs .040/.033). Controlled reranking is therefore fragile; realistic retrieval is the decisive test.

\section{Retrieval Coverage Is the Bottleneck}
In the retrieval-realistic regime, the gold item is not injected. Recall@200 is coverage: did the retriever surface the target at all? Table~\ref{tab:coverage} shows that standard single retrievers cover only 4.6--22.9\% of cases. Strong dense encoders and a two-tower retriever do not remove the bottleneck: best dense coverage remains at most 10.6\%, and the two-tower model does not surpass the strongest CF retriever on CF-strong domains.

\begin{table}[t]
\centering
\caption{Retrieval-realistic coverage@200. Standard single retrievers cover 4.6--22.9\%. Strong dense encoders and a two-tower retriever narrow but do not close the gap; \sys raises coverage to 6.1--24.3\%.}
\label{tab:coverage}
\small
\begin{tabular}{lcccccc}
\toprule
Domain & item-new share & Best standard & Best dense & Two-tower & \sys & Oracle union \\
\midrule
Arts & .354 & .046 & .038 & .034 & .061 & .133 \\
VideoG & .696 & .047 & .046 & .038 & .071 & .144 \\
MIND & .906 & .046 & .050 & .032 & .075 & .094 \\
ML-20M & .642 & .138 & .051 & .107 & .147 & .272 \\
Yelp & .315 & .229 & .106 & .147 & .243 & .522 \\
\bottomrule
\end{tabular}
\end{table}

The reason is structural. Under temporal splits, 32--91\% of test targets are \texttt{item\_new}; these items have zero training interactions. CF and co-occurrence retrievers have no item embedding or history to exploit. Text can help, but even text retrievers are bounded by weak user-to-item semantic matching over a large catalogue. MIND illustrates the extreme: 90.6\% of news targets are new; the best text retriever reaches .046 coverage, 5x the best CF retriever, but still covers only about one in twenty users.

\section{Mitigation: Learned Hybrid Fusion}
\subsection{Oracle headroom and heuristic fusion}
\begin{figure}[t]
    \centering
    \includegraphics[width=0.96\textwidth]{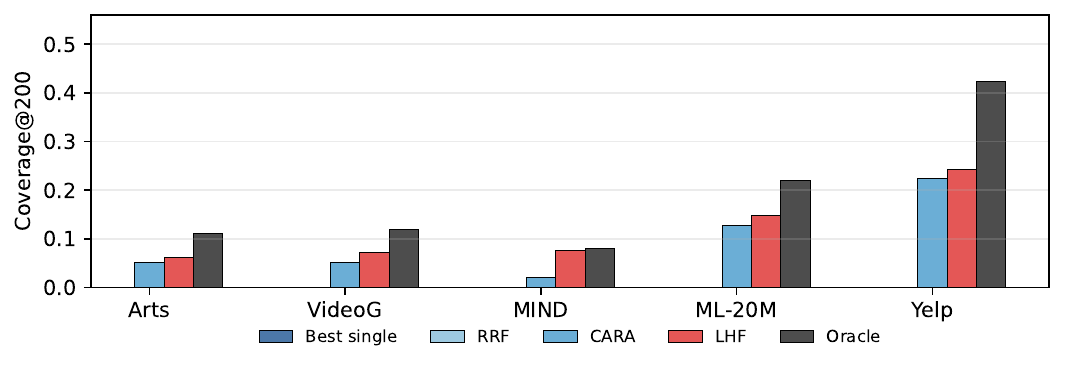}
    \Description{Grouped bar chart comparing best single retriever, RRF, CARA, LHF, and oracle union coverage across five domains.}
    \caption{Retriever complementarity is large, but difficult to realize. \sys is the only combiner that beats every single retriever on all five domains, yet it still captures only part of the oracle union.}
    \label{fig:headroom}
\end{figure}

Different retrievers find different items. The oracle union of standard retrievers raises coverage substantially, yet real single-stage fusions capture little of that headroom. We compare reciprocal-rank fusion (RRF), round-robin interleaving, and a cold-aware retrieval allocation baseline (CARA) that assigns more content-retrieval budget to user-cold requests and more CF budget to warm requests. Figure~\ref{fig:headroom} shows that \sys is the only tested combiner that consistently improves on the best single retriever.

\subsection{LHF: validation-trained fusion over retriever evidence}
\sys scores the union of ten retrievers' top-200 pools. For each user-item pair it uses retriever ranks, scores, present flags, hit counts, min/mean rank, user history length, user-cold flag, item popularity, item-new flag, and item text length. A domain-specific gradient-boosted classifier is trained on the validation split only: positives are validation gold items when covered by the union, and negatives are other union candidates for the same user. At test time, \sys reranks the union and returns top-200. All \sys features are available at serving time and computed from the training snapshot, catalogue metadata, or request-time retriever outputs. The downstream LightGBM ranker used later is also trained only on validation examples and evaluated on held-out test users; no test interactions are used to train either stage.

\begin{table}[t]
\centering
\caption{Serving-time availability of \sys features. All features are computed without validation/test interactions for the target event.}
\label{tab:lhf_features}
\scriptsize
\begin{tabular}{p{.39\columnwidth}p{.52\columnwidth}}
\toprule
Feature group & Source and time boundary \\ 
\midrule
Retriever ranks/scores/present flags & Produced by retrievers from the training snapshot and current request. \\ 
Hit count, min/mean rank & Aggregates over retriever outputs for the same request. \\ 
User history length, user-cold & User interactions observed in the training prefix. \\ 
Item popularity, item-new/cold & Counts from the training prefix only; item-new means zero training interactions at retrieval time. \\ 
Item text length/metadata & Static catalogue metadata available before the request/ranking timestamp. \\ 
\bottomrule
\end{tabular}
\end{table}

\begin{table}[t]
\centering
\caption{LHF procedure for candidate generation.}
\label{tab:lhf_algorithm}
\scriptsize
\begin{tabular}{p{.94\columnwidth}}
\toprule
\textbf{Input:} retriever bank $\mathcal{R}$, budget $N$, validation split. \\ 
1. For each user $u$, form $\mathcal{C}_u=\cup_{r\in\mathcal{R}}\mathrm{TopN}_r(u)$. \\ 
2. Extract retriever evidence and serving-time metadata features $\phi(u,i)$ for each $i\in\mathcal{C}_u$. \\ 
3. Train a domain-specific GBDT classifier on validation positives and in-pool negatives. \\ 
4. At test time, score $\hat{s}(u,i)=f_\theta(\phi(u,i))$ and return top-$N$. \\ 
\textbf{Output:} fused candidate pool for downstream ranking. \\ 
\bottomrule
\end{tabular}
\end{table}

\sys improves coverage on all domains: MIND .046 to .075, VideoG .047 to .071, Arts .046 to .061, ML-20M .138 to .147, and Yelp .229 to .243. It realizes 61\% of oracle headroom on MIND, 25\% on Video Games, and 17\% on Arts, but only 5--7\% on the collaboratively strong domains. The gain is not simply a larger union. Removing cold-start metadata collapses MIND from .075 to .035, below the best single retriever; removing text retrievers collapses MIND further to .019. Logistic regression performs similarly to GBDT, suggesting that the critical signal is the feature set rather than model class.

\begin{figure}[t]
    \centering
    \includegraphics[width=0.96\textwidth]{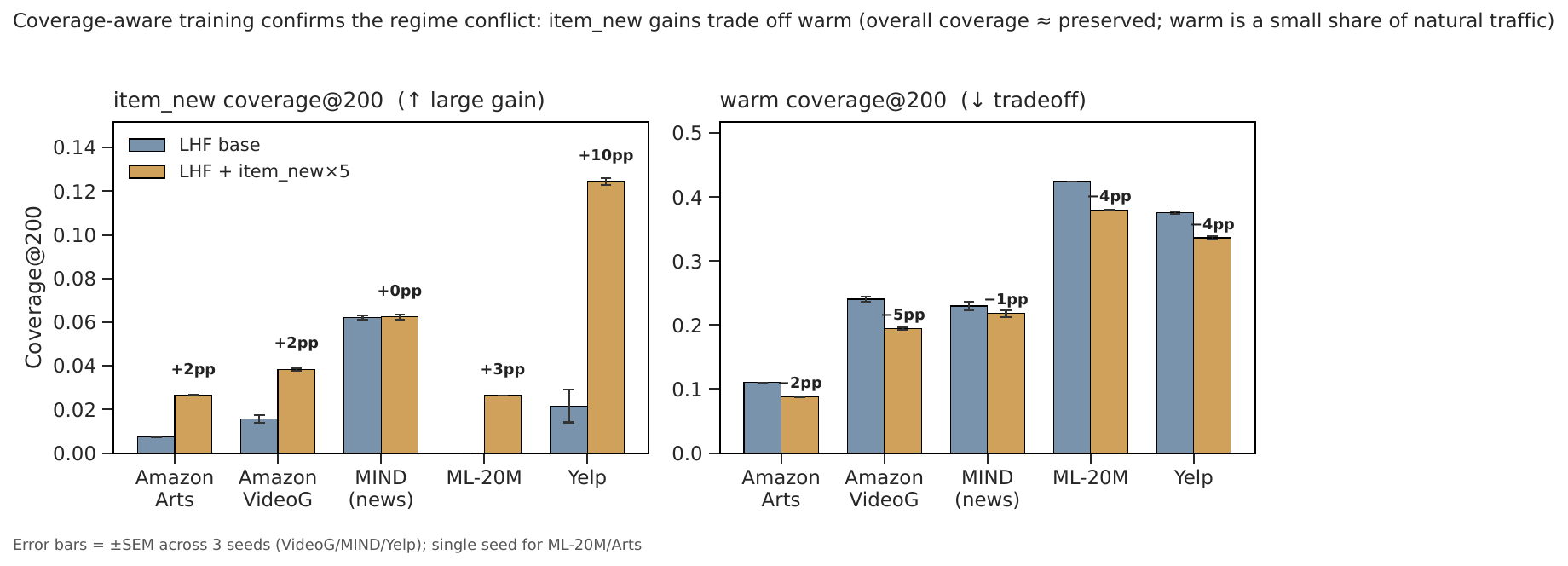}
    \Description{Two-panel bar chart showing that item-new upweighting increases item-new coverage but decreases warm coverage.}
    \caption{Coverage-aware training exposes a regime conflict. Upweighting item-new positives improves new-item coverage, especially on Yelp and Video Games, but trades off warm coverage.}
    \label{fig:coverage_aware}
\end{figure}

Coverage-aware training further confirms the regime conflict (Fig.~\ref{fig:coverage_aware}). Upweighting item-new positives by $5\times$ raises item-new coverage on Yelp by about 10 percentage points and on Video Games by about 2 points, while reducing warm coverage by 4--5 points. The method therefore exposes a tunable allocation tradeoff: cold-start retrieval can be improved, but not for free.

\section{End-to-End: Better Pools Still Need the Right Ranker}
\subsection{Retriever pool to LLM reranker}
Feeding realistic pools to the LLM gives end-to-end Recall@10 at or below 1.4\% across domains. Coverage caps the pipeline, and where coverage is non-trivial, LLM reranking removes more correct top-10 items than it adds. Expanded 1,000-user paired tests confirm this: Yelp LightGCN drops by 3.1 points (37 losses, 6 gains, McNemar $p<.001$), VideoG fusion drops by 1.2 points ($p<.001$), and ML-20M fusion drops by 5.7 points on the stratified subset.

\subsection{The LHF pool contains rankable signal, but not for the LLM}
\begin{figure}[!t]
    \centering
    \includegraphics[width=0.99\textwidth]{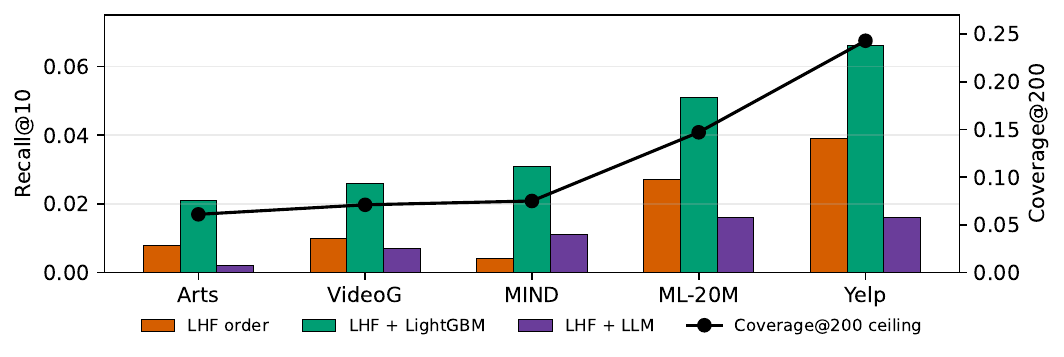}
    \Description{Grouped bar chart comparing LHF-only, LHF followed by LightGBM, LHF followed by LLM, and the coverage ceiling.}
    \caption{The \sys pool contains rankable signal, but the prompt-level LLM does not exploit it. Coverage@200 is the ceiling; a learned LightGBM ranker improves top-10 recall, while LHF$\rightarrow$LLM remains low and often degrades.}
    \label{fig:downstream}
\end{figure}

A key question is whether the pool is bad or the prompt-level LLM is the wrong ranker. We compare three rankers on the same \sys top-200 pool and the same 1,000 LLM-scored users: \sys's own order, a validation-trained LightGBM downstream ranker, and the LLM reranker. Figure~\ref{fig:downstream} shows the outcome. LightGBM improves Recall@10 over \sys-only by 1.3--2.7 percentage points on all five domains (paired $p<.001$ each). The LLM does not: it significantly degrades Yelp and Arts, is not significantly better on the remaining domains, and loses to LightGBM head-to-head on every domain.

\FloatBarrier

This localizes the failure. The problem is not that all realistic pools are useless: a lightweight learned ranker can exploit them. It is also not that semantic cold-start signal is absent: positive-controlled examples and MIND results show that LLMs can rank new content when it is present. The failure is that prompt-level LLM reranking is deployed too late and conflicts with behavioral signals when retrieval is strong.

\section{Auditing Prompt-Level Side Signals}
A natural mitigation is to add graph evidence to the prompt. We audit an internal graph-evidence prompt variant, which we call GraphPrompt+Tail: it includes LightGCN percentiles, co-occurrence evidence, and an additive tail bonus. Table~\ref{tab:gala} decomposes its effect. Graph evidence itself is non-positive on four of five domains; almost all apparent gain comes from the tail prior. Under natural reweighting, that gain disappears except marginally on MIND and can collapse the warm cell.

\begin{table}[t]
\centering
\caption{GraphPrompt+Tail decomposition. Prompt graph evidence rarely helps; apparent gains are mostly tail-prior effects on stratified subsets.}
\label{tab:gala}
\smalltab
\begin{tabular}{lrr}
\toprule
Domain & Evidence effect & Tail-prior effect \\
\midrule
Arts & -.065 & +.170 \\
VideoG & -.017 & +.208 \\
MIND & +.015 & +.023 \\
ML-20M & -.148 & +.010 \\
Yelp & -.041 & +.159 \\
\bottomrule
\end{tabular}
\end{table}

This supports the broader diagnosis: prompt-level side-signal injection can create convincing stratified gains without improving natural traffic. Retrieval-side learning is the more promising place to use semantic or cold-start evidence.

\section{Cost, Discussion, and Limitations}
LLM reranking is expensive. We measure all systems on the same Amazon Video Games candidate pairs on a single H100-class GPU where applicable. Qwen3-8B uses bf16 and the same prompt template, history window, and candidate-text truncation as the accuracy runs; throughput includes prompt/tokenization overhead and forward scoring in our pipeline. LightGCN scores about 55k pairs/s with 4.2GB peak GPU memory; Qwen3-8B scores 58.7 pairs/s and uses 36GB; GraphPrompt+Tail is slower still. Thus the model that does not improve realistic end-to-end recall is also roughly 1000x slower than a CF retriever and about 9x more memory-intensive.

The practical implication is not that LLMs are useless for recommendation. They are useful where text is the only signal, and their semantic judgments can inform offline supervision, item representations, or candidate generation. But spending LLM capacity online at the reranking stage is poorly matched to the dominant failure mode. Future work should target retrieval coverage, learned candidate generation, and lightweight downstream ranking rather than prompt-only reranking.

This study has limitations. LLM reranking is evaluated on sampled subsets for cost reasons, though paired tests and 1,000-user expanded runs reduce this concern. LHF and the downstream LightGBM ranker are both trained on validation data and tested on held-out interactions; we do not claim a nested production training protocol, only a leakage-free offline test of whether the candidate pool contains rankable signal. We leave cross-fitted deployment calibration to future work. We evaluate several prompts but do not exhaustively search prompt designs. Qwen3 8B to 32B is the clean scale axis; Llama-70B is a cross-family 4-bit check. Finally, \sys is a lightweight retrieval-side mitigation rather than a fully optimized production candidate generator. Stronger LLM-distilled or two-tower retrievers may narrow the bottleneck, but our two-tower experiment and item-new analysis suggest they will not remove it without addressing new-item coverage directly.

\section{Conclusion}
LLMs exhibit pockets of semantic cold-start advantage, especially in text-rich domains when the correct item is already in the pool. In realistic retrieve-then-rerank pipelines, that advantage is largely unreachable. Standard retrievers cover the target in only 4.6--22.9\% of top-200 pools, because many targets are genuinely new items with no interaction history. \sys partially mitigates this by learning when to trust different retrievers and cold-start metadata, especially on content-rich domains. Yet even improved pools are better exploited by a lightweight learned ranker than by prompt-level LLM reranking. The cold-start opportunity therefore lives at retrieval and candidate generation, not at the final LLM reranking stage.

\section*{Data and Code Availability}
The reproducibility artifacts for this preprint are archived in two DOI-backed records. The data artifact, available at \url{https://doi.org/10.5281/zenodo.20991039}, contains the processed benchmark material used to reproduce the paper's analyses, including temporal splits, candidate pools, prompts or prompt metadata, model outputs, aggregate metrics, and diagnostic tables where redistribution is permitted by the underlying dataset licenses. The code artifact, available at \url{https://doi.org/10.5281/zenodo.20993306}, contains the evaluation and analysis code for the dual-regime benchmark, retrieval-realistic coverage evaluation, LHF candidate generation, downstream ranking, LLM reranking, plotting, and statistical tests. The corresponding Git repository tag is \url{https://github.com/dongzhe1/rec-diag-bench/tree/0.0.0}.
\FloatBarrier

\clearpage
\appendix
\section{Additional Diagnostic Results}
\subsection{Full positive-controlled reranking table}
Table~\ref{tab:full_rq1} reports the full set of positive-controlled Recall@10 baselines. Confidence intervals are 95\% bootstrap intervals over users. These are the numbers behind the condensed reranking table in the main paper; the conclusion is unchanged: Qwen3-8B is competitive only in some cold/content-heavy cells and never wins a full domain.

\begin{table}[t]
\centering
\caption{Full positive-controlled Recall@10 with 95\% bootstrap intervals.}
\label{tab:full_rq1}
\scriptsize
\begin{tabular}{lccccc}
\toprule
Model & Arts & VideoG & MIND & ML-20M & Yelp \\
\midrule
Popularity & .115 [.09,.15] & .140 [.11,.18] & .038 [.02,.06] & .419 [.35,.49] & .093 [.07,.12] \\
Item-kNN & .098 [.07,.13] & .096 [.06,.13] & .107 [.07,.15] & .371 [.30,.44] & .157 [.12,.19] \\
TF-IDF & .188 [.15,.23] & .132 [.10,.17] & .096 [.06,.13] & .124 [.09,.17] & .113 [.08,.14] \\
Markov & .080 [.06,.11] & .096 [.06,.13] & .057 [.03,.09] & .352 [.29,.42] & .077 [.05,.11] \\
BPR & .083 [.06,.11] & .101 [.07,.13] & .050 [.03,.08] & .486 [.41,.55] & .129 [.10,.16] \\
LightGCN & .147 [.11,.18] & .140 [.11,.18] & .069 [.04,.10] & \textbf{.490 [.42,.56]} & .159 [.12,.20] \\
SASRec & .072 [.05,.10] & .065 [.04,.09] & .042 [.02,.07] & .229 [.17,.29] & .057 [.04,.08] \\
SBERT & .163 [.13,.20] & .107 [.08,.14] & .092 [.06,.13] & .095 [.06,.13] & .082 [.06,.11] \\
Graph-aware & \textbf{.220 [.18,.26]} & .157 [.12,.20] & .084 [.05,.12] & .390 [.32,.45] & \textbf{.159 [.12,.20]} \\
Cross-encoder & .152 [.12,.19] & \textbf{.163 [.13,.20]} & .100 [.07,.14] & .129 [.09,.18] & .085 [.06,.11] \\
LLM-8B & .145 [.11,.18] & .143 [.11,.18] & .092 [.06,.13] & .219 [.17,.28] & .085 [.06,.11] \\
\bottomrule
\end{tabular}
\end{table}

\subsection{End-to-end retrieval and LLM reranking}
Table~\ref{tab:full_e2e} shows the full end-to-end table over the LLM-scored reranking subsets. These subsets are used to measure the costly LLM stage and are not the same denominator as the full-test coverage table; the full-test coverage cap is reported separately in the main paper. In almost every cell, adding the LLM fails to improve top-10 recall; apparent improvements are one-user effects under very low coverage.

\begin{table}[t]
\centering
\caption{End-to-end retrieve-then-rerank on the LLM-scored subset. Coverage is gold-in-pool at top 200; retriever-only and +LLM are end-to-end Recall@10. These subset coverages are not directly comparable to the full-test coverage in the main paper.}
\label{tab:full_e2e}
\scriptsize
\begin{tabular}{llccc}
\toprule
Domain & Retriever & Coverage@200 & Retriever-only R@10 & +LLM R@10 \\
\midrule
Arts & LightGCN & .030 & .007 & .005 \\
Arts & SBERT & .050 & .013 & .007 \\
Arts & RRF & .045 & .013 & .000 \\
Arts & CARA & .048 & .010 & .007 \\
VideoG & LightGCN & .070 & .017 & .000 \\
VideoG & SBERT & .051 & .000 & .000 \\
VideoG & RRF & .070 & .017 & .000 \\
VideoG & CARA & .079 & .011 & .000 \\
MIND & LightGCN & .046 & .004 & .000 \\
MIND & SBERT & .035 & .004 & .000 \\
MIND & RRF & .042 & .008 & .008 \\
MIND & CARA & .069 & .004 & .008 \\
ML-20M & LightGCN & .271 & .057 & .014 \\
ML-20M & SBERT & .024 & .000 & .005 \\
ML-20M & RRF & .271 & .057 & .000 \\
ML-20M & CARA & .257 & .038 & .014 \\
Yelp & LightGCN & .188 & .028 & .013 \\
Yelp & SBERT & .087 & .010 & .003 \\
Yelp & RRF & .162 & .015 & .010 \\
Yelp & CARA & .159 & .018 & .010 \\
\bottomrule
\end{tabular}
\end{table}

\subsection{LHF ablations, coverage-aware training, and cost}
Table~\ref{tab:lhf_ablate} summarizes the ablations most relevant to the method claim. Removing metadata or text retrievers damages the cold-start regimes, while item-new reweighting trades warm performance for new-item coverage. Table~\ref{tab:cost} reports serving cost.

\begin{table}[t]
\centering
\caption{Selected LHF ablations. Values are Coverage@200.}
\label{tab:lhf_ablate}
\scriptsize
\begin{tabular}{lccc}
\toprule
Domain & LHF & No metadata & No text retr. \\
\midrule
Arts & .061 & .058 & .054 \\
VideoG & .071 & .062 & .058 \\
MIND & .075 & .035 & .019 \\
ML-20M & .147 & .144 & .149 \\
Yelp & .243 & .235 & .240 \\
\bottomrule
\end{tabular}
\end{table}

\begin{table}[t]
\centering
\caption{Throughput and peak GPU memory on Amazon Video Games.}
\label{tab:cost}
\scriptsize
\begin{tabular}{lrr}
\toprule
Model & Pairs/s & Peak GPU MB \\
\midrule
Popularity & 503{,}440 & 0 \\
LightGCN & 55{,}455 & 4{,}232 \\
SBERT & 14{,}086 & 4{,}232 \\
Cross-encoder & 583 & 5{,}281 \\
Qwen3-8B LLM & 58.7 & 36{,}103 \\
GraphPrompt+Tail & 42.8 & 40{,}244 \\
\bottomrule
\end{tabular}
\end{table}

\subsection{Additional LHF diagnostics}
\begin{figure}[t]
    \centering
    \includegraphics[width=0.94\textwidth]{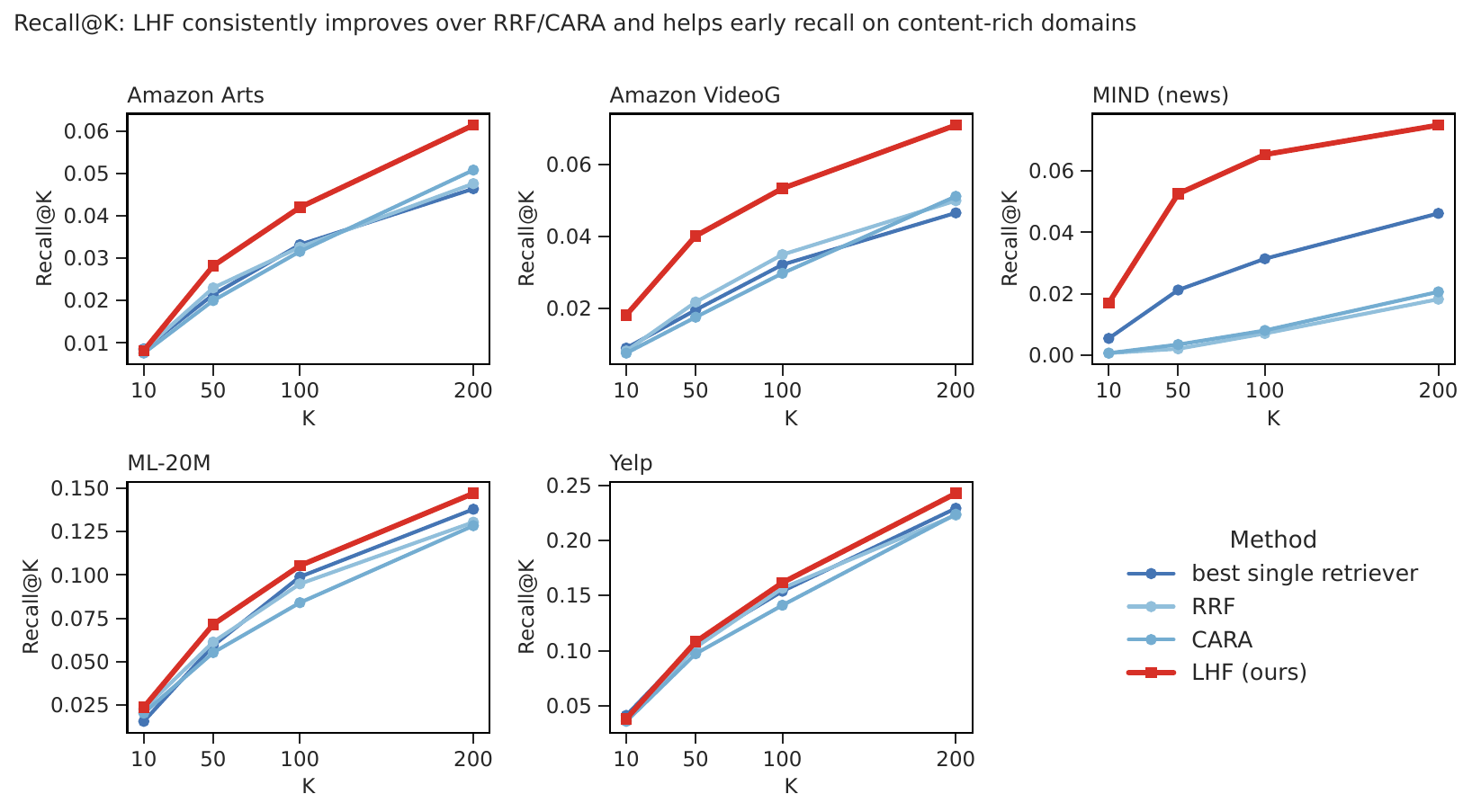}
    \Description{Small-multiple line plots of Recall at different candidate cutoffs for LHF and baselines.}
    \caption{Candidate-generation Recall@K (gold-in-pool coverage at cutoff K) for realizable rankers. \sys beats heuristic fusions across cutoffs and improves early recall on content-rich domains; on CF-strong domains, gains appear mainly at larger K. The oracle union is omitted here because this plot compares executable rankers, not coverage ceilings.}
    \label{fig:recallk}
\end{figure}

Figure~\ref{fig:recallk} compares realized candidate generators across cutoffs; the oracle ceiling is reported separately in the main paper. The learned fusion advantage is not an artifact of the top-200 point. On content-rich domains, \sys improves early recall as well as coverage; on CF-strong domains, the main benefit is coverage at larger candidate budgets, which is why a downstream learned ranker is needed.

\begin{figure}[t]
    \centering
    \includegraphics[width=0.94\textwidth]{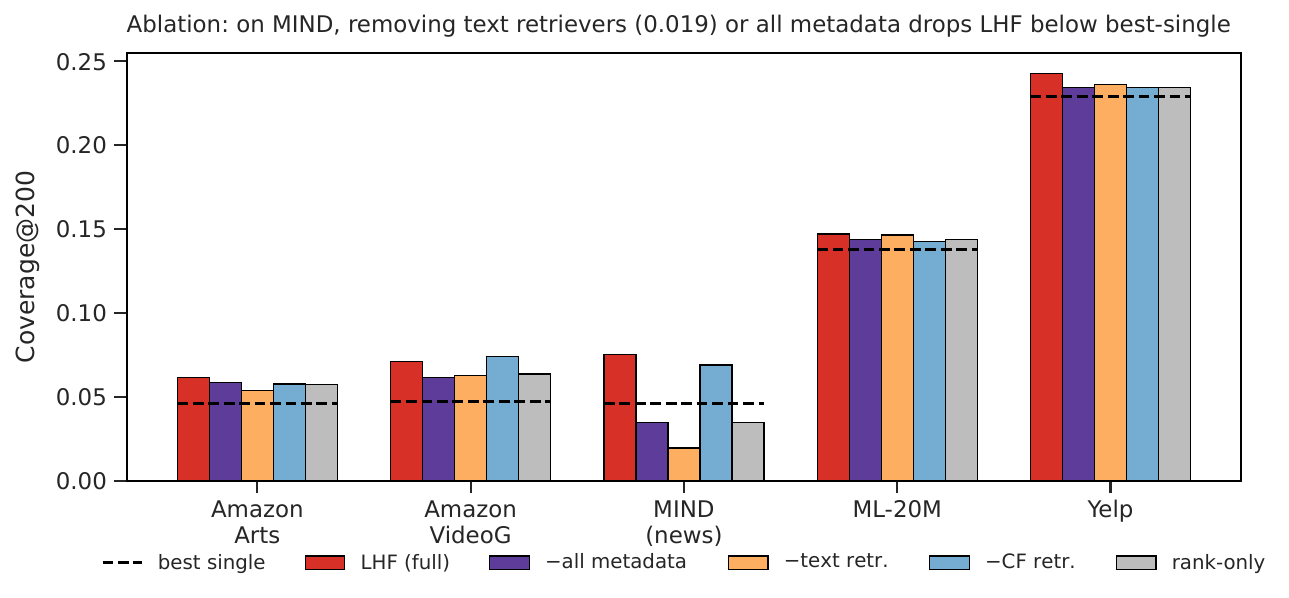}
    \Description{Bar chart comparing LHF ablations by domain and scenario.}
    \caption{LHF ablations. Removing text retrievers or cold-start metadata severely hurts the content-rich cold-start regime, especially MIND. Rank-only features are insufficient without routing metadata.}
    \label{fig:ablation}
\end{figure}

Figure~\ref{fig:ablation} summarizes ablations over the \sys feature set and retriever pool. The largest drops occur when removing text retrievers or cold-start metadata on MIND, confirming that the learned fusion works by routing cold-start cases to the appropriate evidence source rather than merely averaging retrievers.

\begin{table}[t]
\centering
\caption{Strong dense and learned retrieval checks. Best dense encoders and a frozen-encoder two-tower retriever do not eliminate the coverage bottleneck; \sys improves the standard union but remains far below oracle on CF-strong domains.}
\label{tab:dense_full}
\scriptsize
\begin{tabular}{lccccccccc}
\toprule
Domain & Pop. & ItemKNN & TF-IDF & SBERT & LightGCN & Best dense & Two-tower & \sys & Oracle \\
\midrule
Arts & .027 & .019 & .036 & .037 & .046 & .038 & .034 & .061 & .133 \\
VideoG & .042 & .021 & .027 & .035 & .047 & .046 & .038 & .071 & .144 \\
MIND & .003 & .003 & .046 & .040 & .009 & .050 & .032 & .075 & .094 \\
ML-20M & .105 & .138 & .041 & .026 & .131 & .051 & .107 & .147 & .272 \\
Yelp & .187 & .187 & .126 & .096 & .229 & .106 & .147 & .243 & .522 \\
\bottomrule
\end{tabular}
\end{table}

\begin{table}[t]
\centering
\caption{Coverage-aware training over \sys. Item-new upweighting improves new-item coverage with a measurable warm-coverage tradeoff; means and SEM are over three seeds where available.}
\label{tab:regwt}
\scriptsize
\begin{tabular}{lrrrrrr}
\toprule
Domain & LHF overall & LHF item-new & LHF warm & +new$\times$5 overall & +new$\times$5 item-new & +new$\times$5 warm \\
\midrule
Arts & .063 & .007 & .111 & .059 & .027 & .088 \\
VideoG & .074 $\pm$ .002 & .016 $\pm$ .002 & .241 $\pm$ .004 & .079 $\pm$ .001 & .038 $\pm$ .001 & .194 $\pm$ .003 \\
MIND & .076 $\pm$ .001 & .062 $\pm$ .001 & .230 $\pm$ .006 & .075 $\pm$ .001 & .062 $\pm$ .001 & .218 $\pm$ .006 \\
ML-20M & .149 & .000 & .424 & .153 & .026 & .380 \\
Yelp & .242 $\pm$ .002 & .022 $\pm$ .008 & .376 $\pm$ .002 & .235 $\pm$ .002 & .124 $\pm$ .002 & .336 $\pm$ .003 \\
\bottomrule
\end{tabular}
\end{table}

\clearpage
\section*{Ethical Considerations}
This work evaluates offline recommendation datasets and does not deploy a system or collect new user data. The main risks concern privacy in user histories, popularity amplification, unfair long-tail exposure, and overconfidence in LLM-based semantic judgments. Our LLM scoring is performed offline on open-weight models rather than by sending histories to third-party APIs. Any online use of LLM reranking should minimize, truncate, or anonymize user histories before prompting and should audit subgroup coverage and exposure. The public artifacts release processed splits, scripts, prompts, model outputs, and aggregate metrics where redistribution is allowed, while avoiding private user information beyond what is already contained in the underlying public datasets.

\bibliographystyle{unsrtnat}
\bibliography{references}

\end{document}